\documentclass[conference]{IEEEtran}
\IEEEoverridecommandlockouts
\usepackage{graphicx,graphics}
\usepackage{subcaption}
\usepackage{cite}
\usepackage{amsmath,amssymb,amsfonts}
\usepackage{algorithmic}
\usepackage{textcomp}
\usepackage{xcolor}
\usepackage{url}
\def\BibTeX{{\rm B\kern-.05em{\sc i\kern-.025em b}\kern-.08em
    T\kern-.1667em\lower.7ex\hbox{E}\kern-.125emX}}
\begin{document}

\title{The curse of dimensionality: what lies beyond the capabilities of physics-informed neural networks\\
\thanks{The authors acknowledge partial support by the Skoltech program: Skolkovo Institute of Science and
Technology, Russia – Hamad Bin Khalifa University Joint Projects.  }
}

\author{\IEEEauthorblockN{ Javier Penuela}
\IEEEauthorblockA{\textit{Center for Digital Engineering} \\
\textit{Skolkovo Institute of Science and Technology }\\\textit{Moscow, 121205, Russia}\\
javier.penuela@skoltech.ru}
\and
\IEEEauthorblockN{ Henni Ouerdane}
\IEEEauthorblockA{\textit{Center for Digital Engineering}\\
\textit{Skolkovo Institute of Science and Technology }\\\textit{Moscow, 121205, Russia}\\
h.ouerdane@skoltech.ru}

}

\maketitle

\begin{abstract}
Physics-Informed Neural Networks (PINNs) have emerged as a promising framework for solving forward and inverse problems governed by differential equations. However, their reliability when used in ill-posed inverse problems remains poorly understood. In this study, we explore the fundamental limitations of PINNs using a simple illustrative case: RC low-pass filters. Showing that while PINNs can accurately predict system dynamics in forward problems, they fail to recover unique physical parameters when solving inverse problems when more than two parameters are approximated. Our findings provide grounds to understand the boundaries of PINNs applicability for parameter discovery in physical systems.
\end{abstract}


\section{Introduction}
Deep neural networks can be employed as universal approximators for any function if they have enough hidden layers and neurons per layer, and are fed enough data for a given task \cite{HORNIK1989359}. However, as solving efficiently some real-life problems involving, e.g., time series forecasting, remains challenging \cite{CNN}. Hence, the development of many different architectures, as convolutional neural networks, generative artificial intelligence or large language models \cite{CNN, sengar2025generative}. This also applies to modelling systems governed by the laws of physics. To avoid trivial solutions or unstable behaviour, many approaches, including Bayesian neural networks \cite{PENUELA2025113759}, Boltzmann machines \cite{PhysRevB.95.035105}, and Gaussian methods \cite{PhysRevAccelBeams.24.072802} have been used to constrain the optimisation space during the training stage. Recently, a new type of neural network has been gaining traction, especially for solving complex engineering problems. This architecture is called a Physics Informed Neural Network (PINN)\cite{RAISSI2019686,FERNANDEZDELAMATA2023128415}. A PINN is composed of a fully-connected neural network that is regularised using equations that describe the physical laws relevant to a given problem (in the form of the residual of a differential equation (DE)). Usually, the input for a PINN is a set of coordinates (spatial and/or temporal), and the neural network predicts the physical quantity of interest (for example, for the heat equation it would predict the temperature, and for the Navier-Stokes equations, the fluid velocity \cite{RAISSI2019686}). Using autodifferentiation, the mean squared residual of the DE can be calculated and added to the loss function, with any initial and boundary conditions. This results in a regular fully-connected neural network trained using a physics-based regularizer. This architecture has the advantage of allowing to solve the forward problem (FP -- given the observed data, solve the equation) but also the inverse problem (IP -- given the observed data, find the equation constants) \cite{RAISSI2019686}. However, while PINNs have been shown to solve, in some cases, even ill-posed problems, solving the FP is not always feasible.
To the best of our knowledge, there is a poor understanding of the reliability of PINNs when solving ill-posed problems. The main contribution of this work is the exploration of the PINN capabilities and its rarely mentioned limitations, using a simple, illustrative example, focusing on the convergence when tasked with parameter discovery in ill-posed problems. Such limitations arise from the role the neural network plays in the DE solution, which has a fixed dimensionality (in our case, a single dimension --  time) while the problem dimensions increase with the number of parameters to discover in the IP. In section \ref{theory}, we describe our study cases: a single-stage and a two-stage RC low-pass filter, as well as the PINN architecture used. In section \ref{experimental}, the implementation of the experiment is presented, and in section \ref{Results}, we present the results of our modelling experiment. Section \ref{Disscussion}, where we discuss our results, concludes the paper.

\section{Mathematical model} 
 \label{theory}
PINNs have been used to successfully solve different engineering tasks, including heat flow problems, fluid dynamics and other scientific and engineering applications \cite{FERNANDEZDELAMATA2023128415,bdcc6040140}. However, some applications demand better performance and stability, so different variations of the PINN architecture were developed \cite{bdcc6040140}. Moreover, neither the original PINN architecture nor any of its variations has been shown to successfully model circuit behaviour (solve both the FP and IP in relatively simple circuits). The applications of PINNs in circuits are limited to predicting some monotonic characteristics of different devices in power electronics and power systems, like predicting an inverter's frequency response or a transmission state estimation (FP and IP) and solving different difficult problems like optimal power flow in steady-state cases (FP only) \cite{9743327}. In most cases, the used architecture is a variation of PINNs \cite{9743327}. Circuits are a well-suited testbed for PINN capabilities. Especially given the practicality to model different systems using the electrical circuit analogy \cite{hogan2002physical}. For testing the PINN capabilities, we run different tests on two basic circuits: a single-stage and a multiple-stage RC low-pass filter, for the FP and IP, while finding the optimal hyperparameters for each case. The single-stage and two-stage RC low-pass filters are shown in figures \ref{fig:RC1} and \ref{fig:RC2}. In conventional notations, the single-stage filter can be described using Kirchhoff's current law:
\begin{figure}
    \centering
    \begin{subfigure}{0.15\textwidth}
        \centering
        \includegraphics[width=\linewidth]{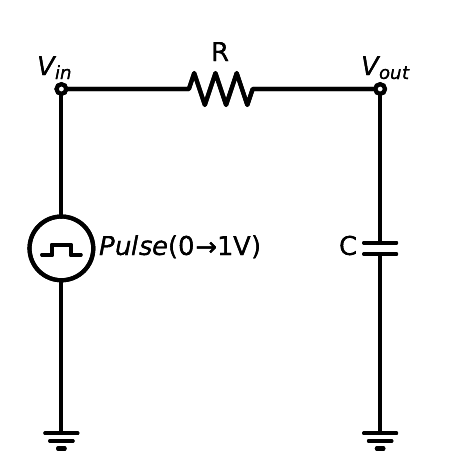}
        \subcaption{\small Single-stage filter}
        \label{fig:RC1}
    \end{subfigure}
    \begin{subfigure}{0.25\textwidth}
        \centering
        \includegraphics[width=\linewidth]{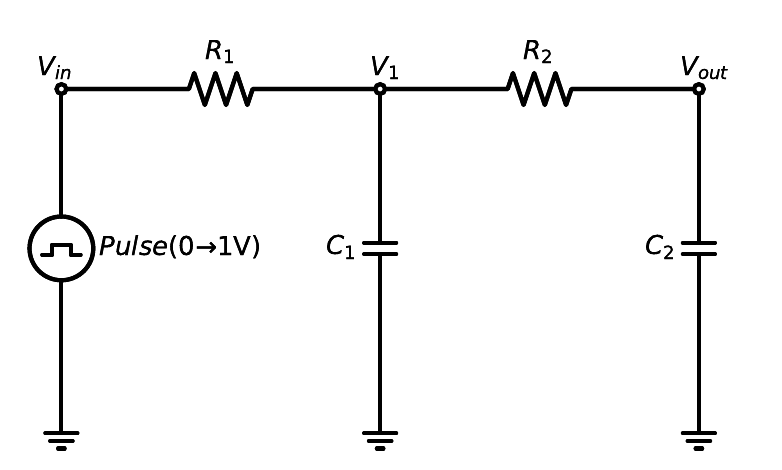}
        \subcaption{\small Two-stage filter}
        \label{fig:RC2}
    \end{subfigure}
    \caption{\small Schematic representation of the simulated low-pass RC filters.}
\end{figure}
\begin{equation}
    i_\mathrm{in}(t)- i_\mathrm{out}(t)=0
    \label{eq:single_node}
\end{equation}
which can be expressed in terms of voltages using Ohm's law and the capacitance:
\begin{equation}
\begin{aligned}
    v_\mathrm{in}(t) - v_\mathrm{out}(t)-RC\frac{dv_\mathrm{out}(t)}{dt}&=0 \label{eq:single_stage_final}
\end{aligned}
\end{equation}
The mean-squared error (MSE) of  (\ref{eq:single_stage_final}) is used for the physics loss for the single-stage case. Note that we may use the circuit representation for other problems, including heat transfer, mass transfer, fluid dynamics and many other problems \cite{hogan2002physical,barooah2008estimation,chen2015electrical} as they satisfy conservation laws. Keeping the generality of the network analogy means that we cannot fix initial or boundary conditions; hence, the IP is, in all cases, ill-posed. Nevertheless, we know that the constants $R$ and $C$ are positive. 

For the two-stage filter, we write Kirchhoff's current law at each node as the following system of equations:
 \begin{equation}
 \left\{
     \begin{aligned}
          i_\mathrm{in}(t)- i_{C_1}(t)-i_2(t) &=0 \\
          i_{2}(t)-i_{C_2}(t)&=0
     \end{aligned}
     \right.
     \label{eq:rc_system} 
 \end{equation}
 and, as for the above case, we obtain 
\begin{subequations}
\label{eq:KCL_two_nodes}
\begin{align}
\frac{v_\mathrm{in}(t) - v_1(t)}{R_1} 
- C_1 \frac{dv_1(t)}{dt} - \frac{v_1(t) - v_\mathrm{out}(t)}{R_2} &=0
\label{eq:KCL_node1} \\[6pt]
\frac{v_1(t) - v_\mathrm{out}(t)}{R_2} 
- C_2 \frac{dv_{\mathrm{out}}(t)}{dt} &=0
\label{eq:KCL_node2}
\end{align}
\end{subequations}
We use the MSE of each (\ref{eq:KCL_two_nodes}) as a variant of the physics loss for the two-node problem. We can combine these equations in two different ways. To obtain a first-order DE, we substitute  (\ref{eq:KCL_node2}) into (\ref{eq:KCL_node1}) to obtain 
 \begin{equation}
\begin{aligned}
v_\mathrm{in}(t) - v_1(t) - R_1C_1 \frac{dv_1(t)}{dt} - R_1C_2 \frac{dv_{out}(t)}{dt}&=0
\end{aligned}
\label{eq:two_node_first_order}
\end{equation}
 The MSE of  (\ref{eq:two_node_first_order}) is also used as a variant for the physics loss for the two-node problem. Equation (\ref{eq:two_node_first_order}) may offer the advantage of reduced complexity. So, we use this formulation to test the behaviour of the IP when partial information of the equation parameters is available. We can also obtain a second-order DE in residual form. Solving (\ref{eq:KCL_node2}) for $v_1$ 
 then we substituting $v_1$ in (\ref{eq:KCL_node1}) we obtain 
\begin{equation}
\begin{aligned}
R_1 C_1 R_2 C_2\,\frac{d^2 v_{\mathrm{out}}(t)}{dt^2}
&+ \big(R_2 C_2 + R_1 (C_1 + C_2)\big)\,\frac{d v_{\mathrm{out}}(t)}{dt}
\\ &+ v_{\mathrm{out}}(t)
- v_{\mathrm{in}}(t)=0. \label{eq:KCL_two_node_second_order}
\end{aligned}
\end{equation}
The MSE of  (\ref{eq:KCL_two_node_second_order}) is used as a variant for the physics loss for the two-stage problem. The resulting equation has the added advantage of fitting a single signal. This may increase convergence. 

Note that the problem is increasingly ill-posed as the number of nodes grows, as we cannot provide an equally growing set of initial and boundary conditions. This problem can be illustrated using the general solution for an RC low-pass filter with $n$ stages for an input step voltage increase $v_{in}$ (charging only) can be presented as:
\begin{equation}
    v_{out}=V_{in}\cdot(1-\sum^n_{i=1}A_i\cdot e^{t \cdot p_{i}})
\end{equation}
where $A_i$ and $p_i$ are constants defined by the circuit architecture. For the sake of simplicity, we limit ourselves to finding the poles of the solution $p_i$. For the single-stage filter (\ref{eq:single_stage_final}) using the auxiliary equation we find
\begin{equation}
    0 - 1-RCp=0 \ \ ,\ \
    p=-\frac{1}{RC}\label{eq:single_stage_tau}
\end{equation}
and analogously for the two-stage filter, we find
\begin{eqnarray}
\nonumber
p_{1,2} & = & -\frac{(R_2 C_2 + R_1 (C_1 + C_2)}{2R_1 C_1 R_2 C_2}\\
& \pm  & \frac{\sqrt{(R_2 C_2 + R_1 (C_1 + C_2)^2 -4R_1 C_1 R_2 C_2}}{2R_1 C_1 R_2 C_2}
\label{eq:KCL_two_node_tau}
\end{eqnarray}
 Note that the combination of $R_i$ and $C_i$ that produces the expected $p_i$ is not unique, hence, the PINN can approximate $p_i$, but the selected $R_i$ and $C_i$ cannot be approximated simultaneously.
 
 \section{Method}
 \label{experimental}
 %\subsection{Method}
 
\begin{figure}[h]
     \centering
     \includegraphics[width=0.5\linewidth]{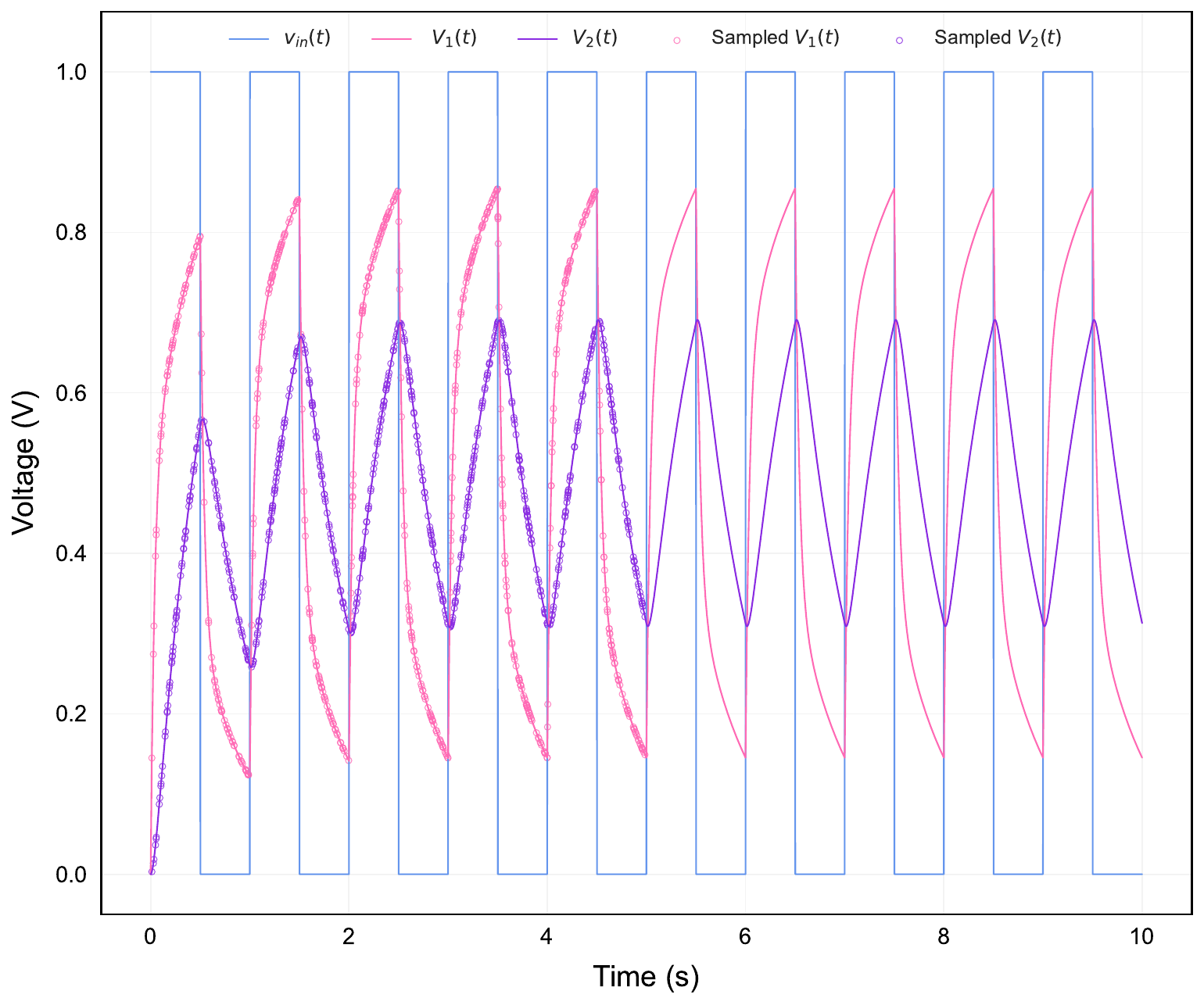}
     \caption{\small Example with 60 samples per cycle from $t=0$ to $t=5$ seconds and the original continuous signal from $t=0$ to $t=10$ seconds for the two-stage RC circuit with the parameters $R_1=15K\Omega$, $C_1=5\mu F$, $R_2=20K\Omega$ and $C_2=15\mu F$.}
     \label{fig:dataset}
 \end{figure}

 \paragraph{Dataset} Using the Scipy \cite{2020SciPy-NMeth} integral solver, we generate a continuous dataset from equations \ref{eq:single_node} and \ref{eq:KCL_two_nodes} with $v_{in}(t)$ formulated as a one-volt step pulse with a one-second period. We check for correctness using a PySpice \cite{PySpice} simulation. To test the influence of data availability on the PINN capabilities to solve ill-posed problems, we use variable sampling density.  The sampling resolution is treated as a hyperparameter. No noise was added to the simulation. A sample of the dataset is presented in Figure \ref{fig:dataset}. 
\paragraph{PINN architecture} We follow the original PINN architecture presented by Raissi et al. \cite{RAISSI2019686}. Similarly to his work, we treat the depth and number of neurons per layer as hyperparameters, with a hyperbolic tangent as the activation function. No special initialisation of the PINN weights is used. The equation parameters for the IP $\mu_i$ (approximation of $C_i$) and $\rho_i$ (approximation of  $R_i$) initialisation, and learning rate are also treated as hyperparameters. For the IP, we use the Swish function \cite{mercioni2021soft} defined as $Swish (X)= X \cdot Sigmoid(X)$ to penalise negative values for the equation parameters. 
The resulting loss function is 
\begin{equation}
\begin{aligned}
   loss=(1-\lambda)\cdot MSE_{data}+\lambda \cdot MSE_{physics} \\+\sum_{1}^{i} Swish(\mu_i) +  \sum_{1}^{i}Swsih(\rho_i) 
   \end{aligned}
\end{equation}
where $\lambda$ is a weight hyperparameter $\leq1$. 
\paragraph{Hyperparameter optimisation}We find the optimal hyperparameters, using the HyperOptSearch optimisation algorithm \cite{bergstra2013hyperopt}, which is shown to outperform random search \cite{NIPS2011}. We use the HyperOptSearch algorithm iteratively. We start by sampling 1000 configurations in a large space search, then using a random forest surrogate model together with one-way partial dependence plots, we narrow down the search space. We repeat this process three times per problem. We evaluate the best fit using the coefficient of determination $R^2$ in  extrapolation in the FP, and the mean absolute percentage error (MAPE) for the poles $p_i$ in the IP. 
 
 \section{Results} 
 \label{Results}
The hyperparameters for the best performing PINN in each problem formulation for extrapolation and parameter estimation are presented in Table \ref{tab:hyperparameters}, and the metrics for the best hyperparameters in all the explored cases are presented in Table \ref{tab:metrics}. For each physics loss, we evaluated three problems:  FP, IP and F/I -- forward problem given the inverse problem PINN structure. 

\begin{figure}
    \centering
    \begin{subfigure}{0.22\textwidth}
        \centering
        \includegraphics[width=\linewidth]{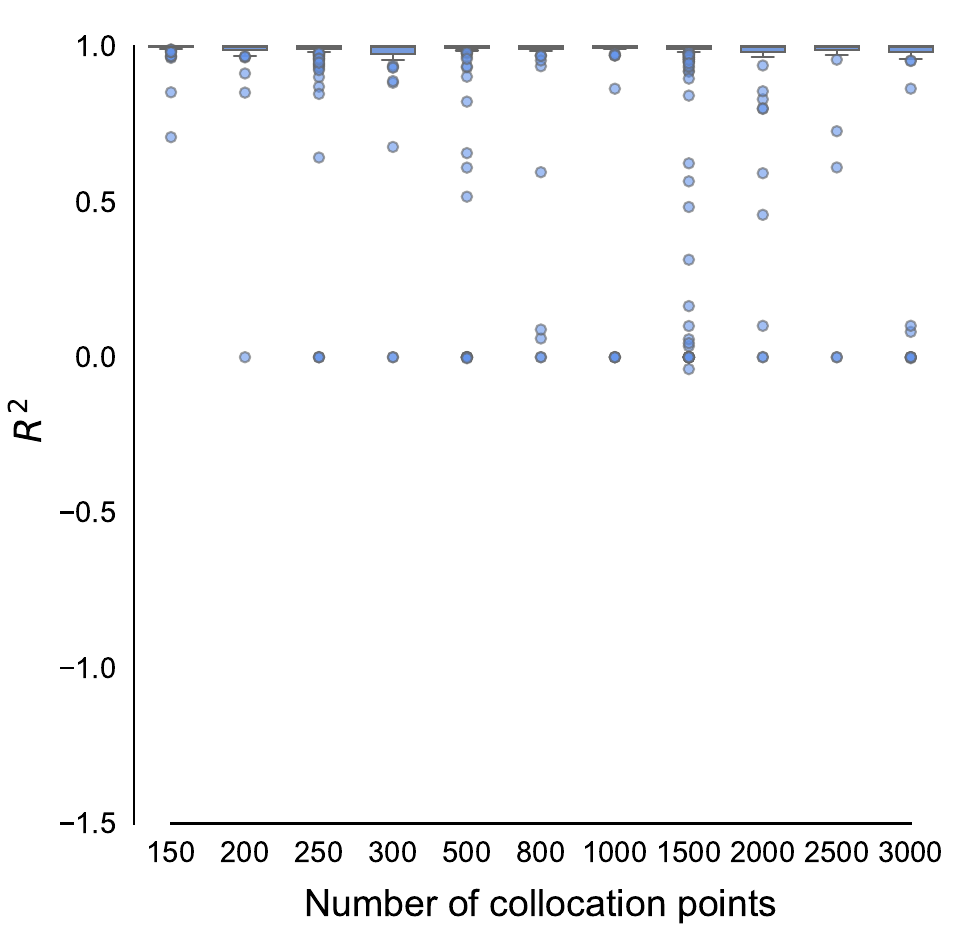}
        \subcaption{\small FP formulation}
        \label{fig:collocation_forward}
    \end{subfigure}
    \begin{subfigure}{0.22\textwidth}
        \centering
        \includegraphics[width=\linewidth]{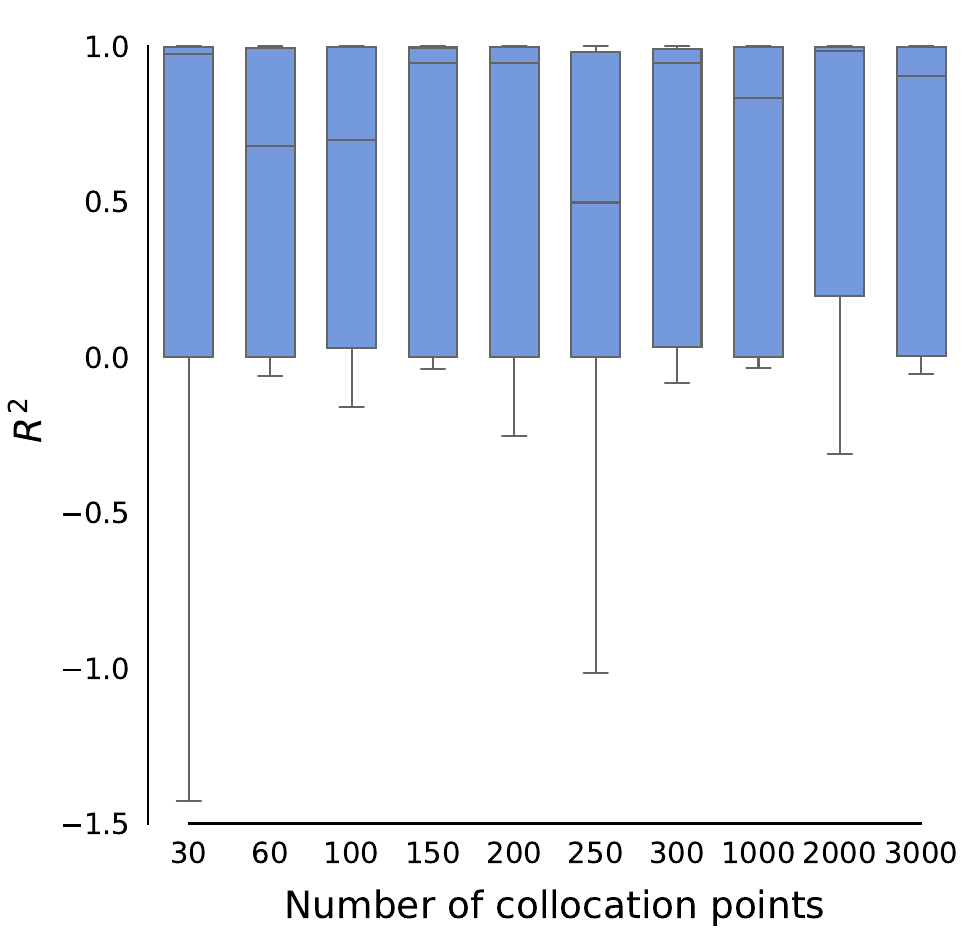}
        \subcaption{\small IP formulation}
        \label{fig:collocation_inverse}
    \end{subfigure}
    \caption{ \small Box plots presenting the distribution $R^2$ against the number of collocation points (C.P.) for the $loss_{phys}$ calculation for the FP after the first iteration of the hyperparameter search.}
    \label{fig:colocation_points}
\end{figure}

During the Hyperparameter selection, we found that the scoring metric variance for the FP formulation is significantly lower than when using the IP formulation (for both FP and IP). An example of the differences is shown in Figure \ref{fig:colocation_points}. Additionally, for the IP, the MAPE tends to stay around 100\%. So the PINN tends to find poles of a few orders of magnitude smaller than the actual values $p_1$ and $p_2$.

The hyperparameters, number of collocation points ($\leq$30), training steps ($\leq$20000) and the learning rate for the equation parameters $\mu$ and $\rho$  do not influence the quality of the resulting model significantly. The initial value of the equation parameters $\mu$ and $\rho$ shows the best performance in the ranges $10e^{-6}\leq\mu\leq 10e^{-3}$ and $10e^{1}\leq\rho\leq 10e^{2}$. Note that the original values for $R_1$ and $R_2$ are not present in the best-performing set of values.
Also we assessed the correlation among $MSE_{loss}$, $MSE_{phys}$, $MSE_{val}$ and $MAPE_{p}$ no stable strong correlation was found.

\begin{table*}
\centering
\footnotesize
\begin{tabular}{|c|c|c|c|c|c|c|c|c|c|c|c|}
\hline
\multicolumn{1}{|c|}{Phys.loss } & 
 &
layers &
neurons &
\begin{tabular}{@{}c@{}}C. P.\end{tabular} &
$\mu$ init &
\begin{tabular}{@{}c@{}}$\rho$ init\end{tabular} &
\begin{tabular}{@{}c@{}}$\lambda$ init\end{tabular} &
\begin{tabular}{@{}c@{}}lr model\end{tabular} &
\begin{tabular}{@{}c@{}}lr $\mu$\end{tabular} &
\begin{tabular}{@{}c@{}}lr $\rho$\end{tabular} &
\multicolumn{1}{c|}{steps} \\ \hline
\multicolumn{1}{|c|}{\raisebox{-3.0ex}[0pt][0pt]{Eq.\ref{eq:single_stage_final}}} & 
FP & 4 & 32 & 150 & -- & -- & 1$e^{-5}$ & 8.258$e^{-3}$ & -- & -- & 19000 \\ \cline{2-12}
\multicolumn{1}{|c|}{} & 
F/I &7 & 124& 100& 1.7$e^{-5}$& 1.29& 1.89$e^{-6}$&7.94$e^{-4}$ & 4.9$e^{-6}$& 2.55$e^{-7}$& 18000\\ \cline{2-12}
\multicolumn{1}{|c|}{} & 
IP & 9&150 &500 &6.8$e^{-4}$ &459.57 &6.84$e^{-8}$ &5.99$e^{-4}$ & 8.76$e^{-6}$& 4.7$e^{-3}$&30000 \\ \hline
\multicolumn{1}{|c|}{\raisebox{-3.0ex}[0pt][0pt]{Eq.\ref{eq:KCL_two_nodes}}} & 
FP & 6& 64&150&-- & --&1.61$e^{-4}$ & 4.77$e^{-3}$ &--  &--  &19000 \\ \cline{2-12}
\multicolumn{1}{|c|}{} & 
F/I &10 &100 & 1000 &4.55$e^{-4}$ & 175
& 3.70$e^{-8}$ & 8.92$e^{-4}$ &4.332$e^{-6}$ &5.55$e^{-7}$ & 45000\\ \cline{2-12}
\multicolumn{1}{|c|}{} & 
IP & 6 & 400 & 150 & 1.43$e^{-5}$ & 294& 6.75$e^{-5}$& 9.92$e^{-4}$ & 3.75$e^{-5}$ &2.26$e^{-7}$& 25000 \\ \hline
\multicolumn{1}{|c|}{\raisebox{-3.0ex}[0pt][0pt]{Eq. \ref{eq:two_node_first_order}}} & 
FP & 9& 32& 150& -- & -- & 6.02$e^{-4}$ & 1.99$e^{-4}$& -- & -- &25000 \\ \cline{2-12}
\multicolumn{1}{|c|}{} & 
F/I & 10& 50& 60& 2.80$e^{-5}$&624 &1.06$e^{-5}$ &9.91$e^{-4}$ &3.90$e^{-4}$ &7.83$e^{-3}$ & 50000 \\ \cline{2-12}
\multicolumn{1}{|c|}{} & 
IP & 16&150 &100 & 1.56$e^{-4}$ &1518 &6.42$e^{-4}$ &9.98$e^{-4}$ &9.19$e^{-7}$ & 1.13$e^{-7}$&  40000\\ \hline
\multicolumn{1}{|c|}{\raisebox{-3.0ex}[0pt][0pt]{Eq. \ref{eq:KCL_two_node_second_order}} }& 
FP &8 & 128& 250& --& --&  6.71$e^{-7}$&9.99$e^{-4}$& --& --& 22000\\ \cline{2-12}
\multicolumn{1}{|c|}{} & 
F/I &7 & 200&1000 & 2.21$e^{-2}$&183 &2.57$e^{-6}$ &8.88$e^{-4}$ &  8.078$e^{-5}$&2.41$e^{-3}$ &  40000\\ \cline{2-12}
\multicolumn{1}{|c|}{} & 
IP &7 &32& 100& 7.77$e^{-2}$& 6149& 4.08$e^{-6}$& 6.75$e^{-4}$& 5.44$e^{-4}$&  1.01$e^{-7}$&12000 \\ \hline
\end{tabular}
\caption{\small Hyperparameters for the best performing models, each equation represents the physics loss for each problem.}
\label{tab:hyperparameters}
\end{table*}

\begin{table*}
\centering
\begin{tabular}{|c|c|c|c|c|c|c|c|c|}
\hline
\multicolumn{1}{|c|}{\begin{tabular}{@{}c@{}}Phys.loss \end{tabular}} & 
& \begin{tabular}{@{}c@{}}ill-posed\end{tabular}&
$MSE_{data}$ &
$MSE_{phys}$ &
$MSE_{val}$ &
$R^2_{data}$ &
$R^2_{val}$ &
$MAPE_{p}$ 
\\ \hline
\multicolumn{1}{|c|}{\raisebox{-3.0ex}[0pt][0pt]{Eq.\ref{eq:single_stage_final}}} &  
FP &no &  1.8$e^{-5}$ & 0.587 & 1.4$e^{-5}$ & 0.99991 & \textbf{0.999905} & -- \\ \cline{2-9}
\multicolumn{1}{|c|}{} & 
F/I & yes& 6.87$e^{-6}$ & 0.426 & 1.1$e^{-5}$&0.999959 &\textbf{0.999928} & 99.7 \\ \cline{2-9}
\multicolumn{1}{|c|}{} & 
IP & yes& 3.93$e^{-4}$ &0.51 & 4.22$e^{-4}$&0.9973 & 0.9972& \textbf{0.17}\\ \hline
\multicolumn{1}{|c|}{\raisebox{-3.0ex}[0pt][0pt]{Eq.\ref{eq:KCL_two_nodes}}} & 
FP & no&4.27$e^{-6}$ &2.25$e^{-9}$ & 6.95$e^{-6}$& 0.999951& \textbf{0.999877} & -- \\ \cline{2-9}
\multicolumn{1}{|c|}{} & 
F/I &yes & 2.042$e^{-7}$ & 1.62$e^{-5}$&4.17$e^{-7}$ & 0.9999975& \textbf{0.9999926}& 82.3/93.4 \\ \cline{2-9}
\multicolumn{1}{|c|}{} & 
IP &yes & 4.28$e^{-2}$  & 4.22$e^{-6}$& 2.05$e^{-2}$ & 0.540&0.55 & \textbf{2.80/4.75} \\ \hline
\multicolumn{1}{|c|}{\raisebox{-3.0ex}[0pt][0pt]{Eq. \ref{eq:two_node_first_order}}} & 
FP & no&  1.59$e^{-6}$ &  2.32$e^{-9}$&  1.17$e^{-6}$& 0.999975&\textbf{0.999969} & -- \\ \cline{2-9}
\multicolumn{1}{|c|}{} & 
F/I &yes & 3.00$e^{-7}$ & 1.30$e^{-3}$&5.31$e^{-7}$ &0.9999961 &\textbf{0.999990} &3.56$e^{9}$/1.50$e^9$  \\ \cline{2-9}
\multicolumn{1}{|c|}{} & 
IP & yes&  7.39$e^{-6}$ &1.62$e^{-7}$ &1.46$e^{-5}$ &  0.999918& 0.99953& \textbf{3.95/ 6.74} \\ \hline
\multicolumn{1}{|c|}{\raisebox{-3.0ex}[0pt][0pt]{Eq. \ref{eq:KCL_two_node_second_order}} }& 
FP & no& 6.75$e^{-7}$ & 0.521&6.92$e^{-7}$
 & 0.999963& \textbf{0.999964}&--  \\ \cline{2-9}
\multicolumn{1}{|c|}{} & 
F/I & yes& 3.40$e^{-7}$ &2.84$e^{-1}$ & 4.023$e^{-7}$&0.999982 & \textbf{0.999978} & 96.01/90.42 \\ \cline{2-9}
\multicolumn{1}{|c|}{} & 
IP & yes& 7.28$e^{-3}$&3.22$e^{-1}$ & 7.68$e^{-3}$& 0.611& 0.582 &  \textbf{52.81/ 0.10} \\ \hline
\end{tabular}
\caption{\small Metrics of the best performing models. For Eq. \ref{eq:KCL_two_nodes},\ref{eq:two_node_first_order} and \ref{eq:KCL_two_node_second_order} MAPE is presented for $p_1/p_2$. }
\label{tab:metrics}
\end{table*}

 \section{Discussion and Conclusions } 
 \label{Disscussion}
 PINNs have been shown to solve ill-posed IP in different applications \cite{9743327}. Using a simple system, we explored the limitations of these applications. From a practical point of view, one would solve the inverse equation in two cases. The first being low-quality and scarce data, one could use the IP to find deviations from the expected parameters and quantify data noise or other external influences. This case is often well-posed, and the performance of the model is fully quantifiable. The second case is when the parameters for a DE are unknown: one could approximate the parameters of the governing DE using a PINN. This case, on which we focus, is often ill-posed. The performance of the model is not fully quantifiable. 
 
 In Figure \ref{fig:colocation_points}, we can observe that for the well-posed FP, the PINN converges easily, in contrast to the ill-posed problem, which shows convergence issues. With certainty, we can state that, as observed from \ref{tab:metrics}, PINNs can indeed solve ill-posed problems. However, the applications of this capability are severely hindered due to instability. Note that although PINNs have low noise sensitivity, real-life data can still increase convergence instability \cite{RAISSI2019686}. We can achieve high accuracy in the validation dataset, and yet, have a trivial or inexact approximation of the DE parameters. Further, we could achieve an accurate parameter estimation and yet have poor predictions. Hence, achieving high-performance predictive capabilities is an insufficient condition to achieve accurate parameter estimation. If the DE parameters are unknown, this limitation renders the use of PINNs unfeasible in most cases.

 As for Eq.~(\ref{eq:single_stage_final}), we could find a well-performing model for the IP, while for Eq.~(\ref{eq:KCL_two_nodes}), we could not. We thus conclude that the $loss_{data}$ plays the role of the initial conditions for the DE. Hence, as we are not solving an ill-posed problem, the advantages of PINNs are reduced to shortcut the formalisation of the initial conditions explicitly for this case. In a more general case, $loss_{data}$ could supplement the boundary conditions in a single dimension. 
 
 We can confirm this interpretation from the Eq.~(\ref{eq:two_node_first_order}) results, where we provided a value for $R_2$, and found an appropriate model for the inverse problem. Using an educated guess is common in several fields, including radio electronics or power electronics, where one would calculate the required filtering frequency threshold; and relying on expert knowledge and available equipment, select a condenser and calculate the corresponding resistance from the pole definition. For similar applications, PINNs could still be applied to discover the unknown parameters for a system. 

\bibliographystyle{ieeetr}
\bibliography{bibliography}

\end{document}